\documentclass[11pt]{article}

\def\la{\langle}
\def\ra{\rangle}
\font\bB=msbm10 at 11pt
\def\bBN{\mbox{\bB N}}
\def\bBC{\mbox{\bB C}}

\def\bBR{\mbox{\bB R}}
\def\1{{\mathchoice{\rm 1\mskip-4mu l}{\rm 1\mskip-4mu l}%
{\rm 1\mskip-4.5mu l}{\rm 1\mskip-5mu l}}}
\def\tr{{\rm Tr}\kern 1pt}
\def\text#1{\quad\mbox{\rm  #1 }\quad}

\newtheorem{lem}{Lemma}
\newtheorem{thm}{Theorem}

\title{A Class of Linear Positive Maps in Matrix Algebras}
\author{A. Kossakowski \\
Institute of Physics \\
Nicolaus Copernicus University \\
Grudziadzka 5, 87--100 Toru\'n, Poland \\
e-mail: kossak@phys.uni.torun.pl}
\date{}

\begin{document}

\maketitle

\begin{abstract}
A class of linear positive, trace preserving maps in $M_n$
is given in terms of affine maps in $\bBR^{n^2-1}$ which map the closed
unit ball into itself.
\end{abstract}

\section{Introduction}

A linear mapping from a $C^\ast$-algebra $A$ into a $C^\ast$-algebra $B$ is
called {\it positive\/} if it carries positive elements of $A$ into
positive elements of $B$. Such a map is said to be {\it normalised\/} if
the image of the unity of $A$ coincides with the unity of $B$.

Positive maps have become a common interest to mathematicians and
physicist, c.f. [1\,--\,31] and [32,\,33]. There are several reasons for
this:
\begin{itemize}
\item[(i)] The notion of positive map generalizes that of state,
$\ast$-homomorphism, Jordan homomorphism, and conditional expectation.
\item[(ii)] By duality, a normalised positive map defines an affine
mapping between sets of states of $C^\ast$-algebras. If these algebras
coincide with the $C^\ast$-algebras of observables assigned to a physical
system, then this affine mapping corresponds to some operation which can be
performed on the system.
\item[(iii)] It has been recently shown that there exists a strong
connection between the classification of the entanglement of
quantum states and the structure of positive maps.
\end{itemize}

Let $M_k(A)$ be the algebra of $k\times k$ matrices with elements from $A$
and $M_k^+(A)$ be the positive cone in $M_k(A)$. For $k\in\bBN$ and
$\varphi:A\to B$ define maps $\varphi_k,\overline{\varphi}_k:M_k(A)\to
M_k(B)$, where $\varphi_k([a_{ij}])=[\varphi(a_{ij})]$ and
$\overline{\varphi}_k([a_{ij}])=[\varphi(a_{ji})]$. The map $\varphi$ is
said to be {\it k-positive\/} ({\it k-copositive}) if the map $\varphi_k$
($\overline{\varphi}_k$) is positive, respectively.

A map $\varphi$ is said to be $k$-{\it decomposable\/} if $\varphi_k$
is positive whenever $[a_{ij}]$ and $[a_{ji}]$ belong to
$M_k^+(A)$. A map $\varphi:A\to B$ is called {\it completely
positive\/}, {\it completely copositive} and {\it decomposable\/}
if it is $k$-positive, $k$-copositive and and $k$-decomposable for
all $k\in\bBN$, respectively. According to the St{\o}rmer theorem
[10], every decomposable map $\varphi:A\to B({\cal H})$ ($B({\cal
H})$ is the set of bounded linear operators on a complex Hilbert
space ${\cal H}$) has the form $\varphi(a)=v^\ast j(a)v$, where
$j:A\to B({\cal K})$ is a Jordan homomorphism and $v$ is a bounded
operator from ${\cal H}$ into ${\cal K}$. Since every Jordan
homomorphism is the sum of a $\ast$-homomorphism and
a $\ast$-antihomomorphism [3] every decomposable map $\varphi:A\to
B({\cal H})$ has the form $\varphi=\varphi_1+\varphi_2$, where
$\varphi_1$ is completely positive and $\varphi_2$ is
completely copositive. A positive map which is not
decomposable is called an {\it indecomposable\/} one.

In the case $A=B({\cal H})=M_n(\bBC)$ every decomposable map
$\varphi:M_n(\bBC)\to M_n(\bBC)$ can be written in the form
\begin{equation}\label{ak1.1}
\varphi(a)\;=\;\sum_ka_k^\ast aa_k+\sum_kb_k^\ast a^Tb_k\,,
\end{equation}
where $a_1,a_2,\ldots\,,$ $b_1,b_2,\ldots\,\in M_n(\bBC)$ and $^T$ is the
transpose map in $M_n$. It is known [2,\,12,\,16] that every positive map
$\varphi:M_2(\bBC)\to M_2(\bBC)$ is decomposable.  The first example of
an indecomposable map in $M_3(\bBC)$ was given by Choi [7,\,8] and its
generalizations can be found in [17\,--\,31].

There are two known types of indecomposable maps in $M_n$ for $n\geq
3$:
\begin{itemize}
\item[(i)] A series of maps $\varphi_k:M_n(\bBC)\to M_n(\bBC)$,
$k=1,\ldots,n-2$; c.f.~[21,\,31]. Let $\epsilon$ be the projection of norm
one in $M_n(\bBC)$ to the diagonal part with respect to a given basis in
$\bBC^n$ and $s$ be the unitary shift in $M_n(\bBC)$ such that
$s=[\delta_{i,j+1}]$, where the indices are understood to be mod $n$. The
maps $\varphi_k$ are defined as follows
\begin{equation}\label{ak1.2}
\varphi_k(a)\;=\;(n-k)\epsilon(a)+\sum_{i=1}^k\epsilon(s^ias^{\ast i})-a\,.
\end{equation}
Since $\varphi_k(\1_n)\;=\;(n-1)\1_n$ and $\tr\varphi_k(a)=(n-1)\tr a$,
the maps
\begin{equation}\label{ak1.3}
\tau_k(a)\;=\;\frac1{n-1}\varphi_k(a)\,,\qquad k=1,\ldots,n-2\,,
\end{equation}
are bistochastic ones.
\item[(ii)] A class of maps $\varphi(p_0,p_1,\ldots,p_n):M_n\to M_n$,
c.f.~[28,\,31] of the form \break
$\varphi(p_0,p_1,\ldots,p_n)(a)=a'$ with
\begin{eqnarray}
a'_{11} &=& p_0a_{11}+p_na_{nn}\nonumber \\
a'_{22} &=& p_0a_{22}+p_1a_{11}\nonumber \\
\cdots && \cdots \nonumber \\
a'_{nn} &=& p_0a_{nn}+p_{n-1}a_{n-1,n-1}\nonumber \\
a'_{ij} &=& -a_{ij}\,,\label{ak1.4}
\end{eqnarray}
where
\begin{equation}\label{ak1.5}
p_0,p_1,\ldots,p_n>0\,,\quad n-1>p_0\geq n-2\,, \quad p_1\cdot\ldots\cdot
p_n\geq (n-1-p_0)^n\,,\quad n\geq 3\,.
\end{equation}
The above maps are atomic, i.e., they cannot be decomposed into the sum of
a 2-positive and a 2-copositive maps [23,\,31].
\end{itemize}

\section{A Class of Positive Maps in $M_n$}
\setcounter{equation}{0}

Let $\varphi$ be a positive, trace preserving map in $M_n(\bBC)$ i.e., a
dynamical map [34]. Since $\varphi$ is self-adjoint, i.e.,
$\varphi(a^\ast)=\varphi(a)^\ast$ it is enough to consider it in the space
\begin{equation}\label{ak2.1}
H_n\;=\;\{a\in M_n\,:\;a=a^\ast\}\,.
\end{equation}
The linear space $H_n$ is the real Hilbert space with the scalar product
$(a,b)=\tr(ab)$ and the norm $\|a\|^2=(a,a)$.

If $a\in H_n$ and $\tr a=p>0$, the necessary condition for positivity of
$a$ is $\tr a^2\leq (\tr a)^2=p^2$, which can also be written in the form
\begin{equation}\label{ak2.2}
\Big\|a-\frac{\1_n}{n}\tr a\Big\|^2\;\leq\;\frac{n-1}{n}(\tr a)^2\,.
\end{equation}
Let us define the following convex sets in $H_n$:
\begin{eqnarray}\label{ak2.3}
B_n(p) &=& \Big\{a\in H_n\,:\;\Big\|a-\frac{\1_n}{n}p\Big\|^2\leq
\frac{n-1}{n}p^2,\quad\tr a=p\Big\}\,,\\
\label{ak2.4}
S_n(p) &=& \Big\{a\in B_n(p)\,:\;a\geq 0\Big\}\,, \\
B_n^0(p) &=& \Big\{a\in H_n\,:\;\Big\|a-\frac{\1_n}{n}p\Big\|^2\leq
\frac{1}{n(n-1)}p^2,\quad\tr a=p\Big\}\,.
\end{eqnarray}
The set $S_n(p=1)$ is just the set of all density operators on $\bBC^n$.
\begin{lem}
If $a\in B_n^0(p)$ then $a\geq 0$, i.e., the inclusions
\begin{equation}\label{ak2.6}
B_n^0(p)\;\subseteq\;S_n(p)\;\subseteq\;B_n(p)
\end{equation}hold.
\end{lem}
{\it Proof.}\quad
Let us observe that $B_n^0$ is invariant with respect to the mapping
$\rho\to U^\ast\rho U$, $U\in {\rm SU}\,(n)$, and the set
$$b_n^0\;=\;\Big\{p_1,\ldots,p_n\,:\;\sum_{i=1}^n\Big(p_i-\frac1np\Big)^2
\;\leq\;\frac{p^2}{n(n-1)}\,,\quad\sum_{i=1}^np_i=p\Big\}
$$
is the ball inscribed in the set
$$
s_n(p)\;=\;\Big\{p_1,\ldots,p_n\,:\;p_i\geq 0\,,\;i=1,\ldots,n\,,\quad
\sum_{i=1}^np_i=p>0\Big\}\,.
$$
Since the eigenvalues of $a\in B_n^0$ lie in $b_n^0$, they are
non-negative.

It should be pointed out that in the case $n=2$ one has
\begin{equation}\label{ak2.7}
B_2^0(p)\;=\;S_2(p)\;=\;B_2(p)\,.
\end{equation}
Let $\varepsilon_p:B_n(p)\to B_n^0(p)$ be defined as follows:
\begin{equation}\label{ak2.8}
\varepsilon_p(a)\;=\;\frac{\1_n}{n}p+\frac{1}{n-1}\Big(a-\frac{\1_n}{n}p
\Big)\;=\;\frac1{n-1}a+\Big(1-\frac{1}{n-1}\Big)\frac{\1_n}{n}p\,.
\end{equation}
The map $\epsilon_p$ is positive and trace preserving. It is convenient to
introduce an orthonormal base in $H_n$: $f_1,\ldots,f_{n^2}$ such that
$(f_\alpha,f_\beta)=\delta_{\alpha\beta}$ and $f_{n^2}=\1_n/\sqrt{n}$,
i.e.~$\tr f_\alpha=0$ for $\alpha=1,\ldots,n^2-1$. Every element of $a\in
H_n$ can be written in the form
\begin{equation}\label{ak2.9}
a\;=\;\sum_{\alpha=1}^{n^2}x_\alpha f_\alpha\,,\qquad x_\alpha\;=\;
\tr(af_\alpha)
\end{equation}or
\begin{equation}\label{ak2.10}
a\;=\;\frac{\1_n}{n}\tr a+\la f,x\ra\,,
\end{equation}where $x=(x_1,\ldots,x_{n^2-1})\in \bBR^{n^2-1}$,
$f=(f_1,\ldots,f_{n^2-1})$, and
\begin{equation}\label{ak2.11}
\la f,x\ra\;=\;\sum_{\alpha=1}^{n^2-1}f_\alpha x_\alpha\,.
\end{equation}
Let $B(\bBR^n,r)$ be the closed ball in $\bBR^n$ with the radius $r$, i.e.,
\begin{equation}\label{ak2.12}
B(\bBR^n,r)\;=\;\{x\in\bBR^n\,:\;\la x,x\ra\leq r\}\,.
\end{equation}
Let ${\cal D}_n$ denote the set of affine maps $(T,b):\bBR^n\to\bBR^n$,
where
\begin{equation}\label{ak2.13}
(T,b)x\;=\;Tx+b\,,\qquad T\in M_n(\bBR)\,,\quad b\in \bBR^n\,,
\end{equation}
which maps the closed unit ball $B(\bBR^n,1)$ into itself. Since ${\cal
D}_n$ is convex, compact and finite-dimensional, each element of ${\cal
D}_n$ can be written as a finite convex combination of its extreme
elements.The extreme elements of ${\cal D}_n$ have been classified in [16]
in thefollowing manner:
\begin{eqnarray}
&\hspace*{-30mm} {\rm Extr}\,{\cal D}_n\;=\;\Big\{(T,b)\,:\;T=R_1\Lambda
R_2\,,\;b=R_1c\,,\;R_1,R_2\in {\rm O}(n)\,,& \nonumber \\
&\Lambda=\|\lambda_i\delta_{ij}\|\,,\;\lambda_1=\ldots=\lambda_{n-1}=
[1-\delta^2(1-\kappa^2)]^{1/2},\;\lambda_n=\kappa[1-\delta^2(1-\kappa^2)]^{
1/2}\,,& \nonumber \\
&c_1=\ldots=c_{n-1}=0\,,\;c_n=\delta(1-\kappa^2)\,,\;0\leq\kappa\leq
1\,,\;0<\delta\leq 1\Big\}\,.& \label{ak2.14}
\end{eqnarray}
It is clear that the map $(T,rb)$ maps the ball $B(\bBR^n,r)$ into itself
provided $(T,y)\in {\cal D}_n$.

Let us observe that the subset Extr${\cal D}_n^0$ of Extr${\cal D}_n$
corresponding to $\kappa=1$ has the form
\begin{equation}\label{ak2.15}
{\rm Extr}\,{\cal D}_n^0\;=\;\{(T,y)\,:\;T\in{\rm O}\,(n)\,,\;y=0\}\,.
\end{equation}
Taking into account (\ref{ak2.3}), (\ref{ak2.9}) and
(\ref{ak2.10}) one finds that an element $a\in B_n(p)$ can be
represented in the form
\begin{equation}\label{ak2.16}
  a\;=\;\frac{\1_n}{n}p+\la f,x\ra\,,\quad p=\tr a\,,
\text{where} x\in
B\Big(\bBR^{n^2-1},p\Big(\frac{n-1}{n}\Big)^{1/2}\Big)\,,
\end{equation}
while the map $\varepsilon_p$, c.f.~(\ref{ak2.8}) takes the form
\begin{equation}\label{ak2.17}
  \varepsilon_p(a)\;=\;\frac{\1_n}{n}p+\frac1{n-1}\la f,x\ra\,.
\end{equation}
Composing the map $\varepsilon_p$ with the map $\psi:B_n(p)\to
B_n(p)$ given by
\begin{equation}\label{ak2.18}
  \psi\Big[T,yp\Big(\frac{n-1}{n}\Big)^{1/2}\Big](a)\;=\;
  \frac{\1_n}{n}p+\Big\la f,yp\Big(\frac{n-1}{n}\Big)^{1/2}\Big\ra
\end{equation}
one finds that the composed map $\varphi[(T,y)]:B_n(p)\to
B_n^0(p)\subseteq B_n(p)$ has the form
\begin{equation}\label{ak2.19}
  \varphi[T,y](a)\;=\;\frac{\1_n}{n}p+\frac1{n-1}\Big\la
  f,Tx+py\Big(\frac{n-1}{n}\Big)^{1/2}\Big\ra\,,
\end{equation}
where $(T,y)\in {\cal D}_{n^2-1}$, is a positive one by the
construction. Since the map (\ref{ak2.19}) is homogeneous it
can be extended to $H_n$ , and its extension to $M_n$ is given by
\begin{equation}\label{ak2.20}
  \varphi[T,y](c+id)\;=\;\varphi[T,y](c)+i\varphi[T,y](d)\text{for}
  c,d \in H_n\,.
\end{equation}
The main result can be summarized in the following
\begin{thm}
Every affine map $(T,y)$ which maps the closed unit ball in
$\bBR^{n^2-1}$ into itself induces a positive, trace preserving map
$\varphi(T,y):M_n\to M_n$
\begin{equation}\label{ak2.21}
  \varphi[T,y](a)\;=\;\frac{\1_n}{n}\tr a+\frac1{n-1}\Big\la f,
  Tx+y\Big(\frac{n-1}{n}\Big)^{1/2}\tr a\Big\ra\,,
\end{equation}
where $x=(x_1,\ldots,x_{n^2-1})\in\bBR^{n^2-1}$,
$x_\alpha=\tr(af_\alpha)$, and $f=(f_1,f_2,\ldots,f_{n^2-1})$,
$f_\alpha=f^\ast_\alpha$, $\tr(f_\alpha
f_\beta)=\delta_{\alpha\beta}$, $\tr f_\alpha=0$,
$\alpha,\beta=1,\ldots,n^2-1$.
\end{thm}
It should be pointed out that in the case $n=2$ every positive,
trace preserving map has the form (\ref{ak2.21}).

A systematic construction of the operators
$f_1,\ldots,f_{n^2-1}\in H_n$ (generators of SU$(n)$) is well
known, c.f.~[35]. They are given by
\begin{eqnarray}
(f_1,\ldots,f_{n^2-1}) &=& (d_\ell,u_{k\ell},v_{k\ell})\,,\qquad
\ell=1,\ldots,n-1\,,\quad 1\leq k<\ell\leq n\,,\qquad\mbox{}\\
d_\ell &=& \frac{1}{\sqrt{\ell(\ell+1)}}\sum_{k=1}^\ell(e_{kk}-
ke_{k+1,k+1})\,,\quad \ell=1,\ldots,n-1\,,\\
u_{k\ell} &=& \frac{1}{\sqrt{2}}(e_{k\ell}-e_{\ell k})\,,\\
v_{k\ell} &=& \frac{-i}{\sqrt{2}}(e_{k\ell}-e_{\ell k})\,,\\
e_{k\ell} &=& e_k(e_\ell,\,\cdot\,)\,,
\end{eqnarray}
where $(e_1,\ldots,e_n)$ is an orthonormal base in $\bBC^n$.

As an illustration of the formula (\ref{ak2.21}) let us consider
the case $n=3$. Using the notation
\begin{equation}\label{ak2.27}
  x_1\;=\;\tr(ad_1)\,,\qquad x_2\;=\;\tr(ad_2)
\end{equation}
and
\begin{equation}\label{ak2.28}
  x_{k\ell}\;=\;\tr(au_{k\ell})\,,\qquad
  y_{k\ell}\;=\;\tr(av_{k\ell})
\end{equation}
let us consider the rotation $R(\alpha)\in\,$SO(8) given by
\begin{eqnarray}
x_1' &=& x_1\cos\alpha-x_2\sin\alpha\,,\nonumber\\
x_2' &=& x_1\sin\alpha+x_2\cos\alpha\,,\nonumber\\
x_{k\ell}' &=& -x_{k\ell}\,,\label{ak2.29} \\
y_{k\ell}' &=& -y_{k\ell}\,,\qquad 1\leq k<\ell \leq n\,.\nonumber
\end{eqnarray}
Taking into account the explicit form of $d_1$, $d_2$,
$u_{k\ell}$, $v_{k\ell}$, $1\leq k<\ell\leq 3$, one finds that the
corresponding map
$\varphi[\alpha]\equiv\varphi[R(\alpha),0]:M_3(\bBC)\to M_3(\bBC)$
has the form $a'=\varphi[\alpha](a)$, where
\begin{eqnarray}
a_{11}' &=&
\frac12[\lambda(\alpha)a_{11}+\mu(\alpha)a_{22}+\nu(\alpha)a_{33}]\,,\nonumber
\\
a_{22}' &=&
\frac12[\nu(\alpha)a_{11}+\lambda(\alpha)a_{22}+\mu(\alpha)a_{33}]\,,\nonumber
\\
a_{33}' &=&
\frac12[\mu(\alpha)a_{11}+\nu(\alpha)a_{22}+\lambda(\alpha)a_{33}]\,,\nonumber
\\
a_{ij}' &=& -\frac12a_{ij}\,,\qquad i\neq j\,,
\end{eqnarray}
and
\begin{eqnarray}
\lambda(\alpha) &=& \frac23(1+\cos\alpha)\,,\nonumber \\
\mu(\alpha) &=& \frac23\Big(1-\frac12\cos\alpha-\frac{\sqrt{3}}{2}\sin\alpha\Big)
\,,\nonumber \\
\nu(\alpha) &=&
\frac23\Big(1-\frac12\cos\alpha+\frac{\sqrt{3}}{2}\sin\alpha\Big)
\,,\nonumber \\
\lambda(\alpha) &+& \mu(\alpha)\;+\;\nu(\alpha)\;=\;2\,.
\end{eqnarray}
The special cases are:
\begin{itemize}
\item[(i)] $\quad a'=\varphi[\alpha=\pi/3](a)$
\begin{eqnarray*}
a_{11}' &=& \frac12(a_{11}+a_{33})\,,\\
a_{22}' &=& \frac12(a_{22}+a_{11})\,,\\
a_{33}' &=& \frac12(a_{33}+a_{22})\,,\\
a_{ij}' &=& -\frac12a_{ij}\,.
\end{eqnarray*}
\item[(ii)] $\quad a'=\varphi[\alpha=-\pi/3](a)$
\begin{eqnarray*}
a_{11}' &=& \frac12(a_{11}+a_{22})\,,\\
a_{22}' &=& \frac12(a_{22}+a_{33})\,,\\
a_{33}' &=& \frac12(a_{33}+a_{11})\,,\\
a_{ij}' &=& -\frac12a_{ij}\,.
\end{eqnarray*}
\item[(iii)] $\quad a'=\varphi[\alpha=\pi](a)=1/2(\1_3\tr a-a)$
\item[(iv)] $\quad
a'=\varphi[\alpha=0](a)=1/3\varphi[\alpha=\pi](a)+(1-1/3)\psi(a)$
\begin{eqnarray*}
a'' &=& \psi(a)\,, \\
a_{11}''&=& a_{11}\,,\quad a_{22}''\;=\;a_{22}\,,\quad
a_{33}''\;=\;a_{33}\,,\\
a_{ij}'' &=& -\frac12a_{ij}\,.
\end{eqnarray*}
\end{itemize}
The maps (i) and (ii) are indecomposable Choi's maps, the map
(iii) is completely copositive, while (iv) is decomposable.

Let $[x_{ij}]\in M_3(M_3)$ be the matrix
\begin{equation}\label{ak2.36}
  [x_{ij}]\;=\;\left(
  \begin{array}{ccc|ccc|ccc}
    1 & 0 & 0 & 0 & 1 & 0 & 0 & 0 & 1 \\
    0 & p & 0 & 0 & 0 & 0 & 0 & 0 & 0 \\
    0 & 0 & p^{-1} & 0 & 0 & 0 & 0 & 0 & 0 \\
    \hline
    0 & 0 & 0 & p^{-1} & 0 & 0 & 0 & 0 & 0 \\
    1 & 0 & 0 & 0 & 1 & 0 & 0 & 0 & 1 \\
    0 & 0 & 0 & 0 & 0 & p & 0 & 0 & 0 \\[0.8mm]
    \hline
    0 & 0 & 0 & 0 & 0 & 0 & p & 0 & 0 \\
    0 & 0 & 0 & 0 & 0 & 0 & 0 & p^{-1} & 0 \\
    1 & 0 & 0 & 0 & 1 & 0 & 0 & 0 & 1
  \end{array}\right)\qquad p>0\,.
\end{equation}
Then both $[x_{ij}]$ and $[x_{ji}]$ belong to $M_3(M_3)^+$. It is
easily seen that the matrix $[\varphi[\alpha](x_{ij})]$ is not
positive for $0<\alpha<\pi$, i.e., the map $\varphi[\alpha]$ is
indecomposable for $0<\alpha<\pi$.

It should be pointed out that the indecomposable maps (\ref{ak1.4}) does
not belong to the class considered above.

\end{document}